\documentclass[final,3p,times,twocolumn]{elsarticle}

 \usepackage{graphicx}

\usepackage{amssymb}
\usepackage{color}
\usepackage{listings}
\usepackage{mdwlist}
\usepackage{algorithm}
\usepackage{algpseudocode} 
\usepackage{url}

\definecolor{mygreen}{rgb}{0,0.6,0}
\definecolor{mygray}{rgb}{0.5,0.5,0.5}
\newtheorem{definition}{Definition}

\begin{document}

\begin{frontmatter}



\title{Component Matching as a Graph Matching Problem}


\author{Suresh Kamath}

\address{Consultant. \\ North Carolina}

\begin{abstract}
The development of an IT strategy and ensuring that it is the best possible one for business is a key problem many organizations face. This problem is that of linking business architecture to IT architecture in general and application architecture specifically. In our earlier work we proposed Category theory as the formal language to unify the business and IT worlds with the ability to represent the concepts and relations between the two in a unified way.  We used rCOS as the underlying model for the specification of interfaces, contracts, and components.  The concept of pseudo-category was then utilized to represent the business and application architecture specifications and the relationships contained within. Contracts are used for the specification of both IT and Business architecture components. The linkages between them is now established using the matching of the business component contracts with the application component contracts. Typically,  the matching was based on manual process, in this paper we extend the work by considering automated component matching process. In this paper we provide implementation of the matching process using graph matching. 
\end{abstract}


\end{frontmatter}


\section{Introduction}
\label{sec:intro}
The development of an IT strategy and ensuring that it is the best possible one for business is a key problem many organizations face. This problem is that of linking business architecture to IT architecture in general and application architecture specifically. In our earlier work  \cite{Kamath:ICISA.2011} ,\cite{Kamath:WICSA.2011.12}, and \cite{IGI:978-1-4666-2199-2} we proposed Category theory as the formal language to unify the business and IT worlds with the ability to represent the concepts and relations between the two in a unified way.  We used rCOS \cite{report406} as the underlying model for the specification of interfaces, contracts, and components.  The concept of pseudo-category \cite{DBLP:journals/jcst/Lu05} was then utilized to represent the business and application architecture specifications and the relationships contained within.  Contracts are used for the specification of both IT and Business architecture components. The linkages between them is now established using the matching of the business component contracts with the application component contracts. Typically,  the matching was based on manual process, \cite{kamath2024componentmatchingapproachlinking}, in this paper , we extend the work by considering automated component matching process. We also provide implementation of the matching process using graph matching. 

The rest of the paper is organized as follows. In Section \ref{sec:prob} we define the problem to be addressed followed by Section \ref{sec:related} wherein we review some of the related work in Component and Contract matching and software adaptability aspects. We discuss the neural approach to our solution in Section \ref{sec:solution} followed by the details in Section \ref{sec:matching}. We provide concluding remarks in Section  \ref{sec:conclusion}.

\section{Problem Statement}
\label{sec:prob}
We have covered the use cases and various definitions used in this section in  \cite{kamath2024componentmatchingapproachlinking}. Given two contracts, $Ctr_1$ and $Ctr_2$, we can now define the notion of refinement, see \cite{report350} for details. Essentially the refinement notion states that $Ctr_2$ are either equivalent or  $Ctr_1 \sqsubseteq Ctr_2$, in which case $Ctr_2$ provides the same services as $Ctr_1$ and is not easier to diverge or to deadlock than $Ctr_1$. Next we need the concept of complete contract, which is defined in \cite{comppublication10}, and means that no execution of a trace in the protocol $Prot$ from an initial state will enter a blocking state.

For example, Listing \ref{contrDM} shows the contract specification for the component (in rCOS) that implements the \verb+DocumentManager+ interface (business).

\lstdefinestyle{java}{
basicstyle=\small,
language=Java, 
rulecolor=\color{black},
breaklines=true,
numbers=left,
numbersep=5pt, 
numberstyle=\tiny\color{black},
commentstyle=\color{black},
keywordstyle=\bfseries\color{black},
stringstyle=\ttfamily
}
\lstset{style=java}

\begin{lstlisting} [caption={rCOS Component DocumentManager},label=contrDM]
contract ctr_DM of  DocumentManager {
  //variables and initialization code
  private Document[] documents;
  private Document document;
  //provided methods
  public void viewDocument(String documentId) {
      //implementation code;
      ;}
  public Document[] searchDocuments(String params) {//implementation code;
  ;}
  public void setPreference(String documentType, String preference) {//implementation code;
  ;}

  //private methods
  private Connection connect() {//code to connect to the repositroy;
  ; }

  //required services
  private Document getDocument(String documentId){;};
  private Document[]  getDocuments(String criteria){;};
  private Boolean updateDocumentSetting(String documentType){;};

  protocol {
      ((searchDocument+?setPreference)*| (searchDocument+ viewDocument?setPreference)*  )
  }
}
\end{lstlisting}
 In our approach we will use the contract specification as the starting point for the matching process and hence the publication aspects of the contract as discussed in \cite{kamath2024componentmatchingapproachlinking} are ignored in the current approach The basic requirements of a contract can be stated as:

\begin{itemize*}
\item The contract specifies which component contract it is describing (contact $<$contract name$>$ of $<$component class$>$
\item Specification of variables used
\item Methods in the contact specified with arguments used, variable definitions within the method. We do not need the implementation code for the methods.
\item For readability purposes, we can group the methods into public, private and required. This is however not used in the matching algorithm.
\item The provided protocol can be expressed using regular expression.
\end{itemize*}

Given the current application architecture components and their contracts and a contract for a new business architecture component, the problem is to search the application architecture contracts and identify component(s) that match the given contract. We expect the match to include (a) match of the variable types, method types and method variable declaration, and (ii) match the protocol specification. For the definition of the application architecture, business architecture, see \cite{kamath2024componentmatchingapproachlinking}.

\section{Brief Review of the Past Work - Component Matching}
\label{sec:related}
We provide a brief overview of the work in software component matching in this section. Rollins and Wing \cite{Rollins91specifications} proposed an approach based $\lambda$Prolog to be used as the query as well as specification language. The specification of a component may consist of name, signature and pre- and post-conditions on the behavior of the component. The requirement (query) is also specified using $\lambda$Prolog and then matching between the query and a set of specifications is performed (\verb+satisfy+ as logical implication). They also propose measurements such precision and recall to narrow the search results (closer to the query satisfaction). The main draw back is the performance involved to do the specification matching, when practical size for the set of components is involved.

Zaremiski and Wing \cite{Zaremski:1995:SMT:210134.210179}, \cite{Zaremski:1996:SSM:923008} consider the problem of efficient signature matching as opposed to the specification matching, which also matches the component behavior. The matching procedures are implemented in ML and statistics on the performance when using a variety of matching procedures are provided. Two types of components are considered, function and module. A function is nothing but a function or method declared in a particular software artifact, for example mathematical functions. The matching is based on the function's type, that is types of both input and output parameters. The module on the other hand is a group of functions - e.g. C++ or Java Class, Ada packages etc. In this case the type is based on the an interface consisting of user defined type and function types. The matching is performed exactly as above but using the interface. The paper provides algorithms for exact match as well as several relaxation matching. 

Zaremski and Wing \cite{Zaremski:1997:SMS:261640.261641} extends their earlier work on signature matching to include the specification matching.  The matching uses Larch/ML interface language to state the pre- and post- conditions and use theorem proving (using Larch) to determine match and mismatch. The matching is extended to function and module as is the case in their previous research. Various "qualities" of matching methods are defined and two application use cases one for for retrieval for reuse and other for subtyping of object-oriented types are presented.

Goguen et. al \cite{DBLP:journals/jsi/GoguenNMLZB96} proposes a component search approach combining the simplicity of key word search for performing faster search with specification-based search for very accurate matching. The concept of ranking is introduced to qualify the query matches. Another idea is the use of multi-level filtering to improve the efficiency of search. The filtering criteria includes semantic filtering. The detailed discussion of these ideas are beyond the scope of this paper and refer the paper cited.

Mili et. al \cite{605762} discusses the design and implementation of storing and retrieving software components from a repository based on formal specifications. The formal specification is called relational specification. The query for retrieval is also a relational specification. The specification is of the form (S,R), where S is a set called space of the specification, nothing but the attributes and variables that will be used in any program on S, and R is a relation on S, a subset of SxS. An ordering between specification is utilized to organize the storage structure of components in the repository which will also allows efficient retrieval using queries. Another interesting aspect of this implementation is that when the system can not find a match, components that come closer to satisfying the query can be retrieved.  

Pahl \cite{DBLP:conf/fm/Pahl01} uses description logic to develop an ontology for matching of components. The description logic consists of three types of entities names objects,concepts and relationships between concepts. The component description and matching ontologies are developed using this. A component is described using the functional behavior and interaction protocol, which together forms the contract for a component. The matching process involves subsumption (relationship involving concepts and roles),  matching of component operation descriptions and matching of component interaction protocols.

There are several other works related to component matching and we provide the references here, Penix and Alexander \cite{DBLP:journals/ase/PenixA99}, Morel and Alexander \cite{10.1109/ASE.2003.1240302} Wang and Krishanan \cite{WangKrishnan06}, Lau et. al \cite{LauNg:12} and related to service matching in web services, Heckel et. al \cite{DBLP:journals/entcs/HeckelCL04}, \cite{DBLP:conf/icalp/EngelsH00}, Iribarne et. al \cite{Iribarne2004} and Iribarne and Vallccillo \cite{IV00}.

In \cite{7320344}, the authors provide a framework to create the specification profile of components using the CBSE ontology, \cite{szyperski2002component}, also \cite{mili2001reuse}. The proposed framework approach consists of profiling of the components using the Extended Backus-Naur Form, which describes the desired properties of the required components followed by an automatic search and retrieval mechanism for finding appropriate components for reuse.

In \cite{CASTILLOBARRERA2023107282}, Castillo-Barrera et.al  provides a framework for verifying the matching of software components that does not require the user possessing highly specialized skills and is able to check contract conformance of functional semantics aspects. They make use of architecture description languages (ADLs) to specify configurations of component interconnections. Interface contracts are specified with a customized version of CORBA-IDL and employ ontology reasoning engines to check conformance among interface contracts.

\section{Component matching as a Graph matching problem}
\label{sec:solution}
Our solution to the component matching problem is based on the following steps:
\begin{enumerate*}
\item Create a graph representation of the contracts. Here each contract as nodes representing the core variables, variable types, methods, method arguments, method variables, and return types. The edges of the graph representing the relationships of these. So, the application architecture can thus be represented as a composite graph compose of a number of contract graphs.
\small\begin{verbatim}
                 AA
                  |
    -------------------------------
    |     |     |     |      |     |  
                             
 comp1 comp2 comp3 comp4 comp5   comp6  
 (other nodes that represent 
      variables, methods etc.)   
\end{verbatim}\normalsize
\item  Represent the new business architecture component also can be represented as a graph, reqGrapg.
\item Look for a subgraph of the application architecture graph that matches the reqdGraph. This is a well known problem in graph theory that has solutions. The case of exact graph matching is known as the graph isomorphism problem and problem of exact matching of a graph to a part of another graph is called subgraph isomorphism problem, which is our case, see Section \ref{sec:subgraphmatch} for further details.
\item We have a final step where we need to match the protocols of the architecture contracts that matched and the required contract. This will be done by converting the protocols into DFAs and looking to see if they are equivalent, see Section \ref{sec:protocolmatch}..
\end{enumerate*}
We need to address several aspects before we can address the solution, namely (a) specifying contracts, (b) representing contracts as graph, (b) how do we create the application architecture graph consisting of many contracts, and finally (c) performing the matching and identifying the matching component(s). The rest of the paper addresses each of these tasks.
 
\subsection{Components and Contracts}
\label{sec:contracts}
In this section we present the slightly modified representation of contract of a component that is used in the implementation.

\begin{definition}
A publication contract or contract of a component $Ctr_{component}$ is a tuple,  is (Q,I,D,T) where:
\begin{itemize*}
\item Q is a set of variable initializations;
\item I is the interface I (provided methods);
\item J is the interface of required methods J
\item D is a function that defines each method m of I with a design (no guard);
\item T  is a protocol (set of traces) over the methods m of I and J.
\end{itemize*}
\end{definition}

We, however relax the restriction of providing the function design and may provide the variables and return values. Java language conventions will be followed for specifying the contarct and a sample can be seen in Listing \ref{contrDM}.

\subsection {Contract Grammar}
\label{sec:grammar}
We have developed the grammar for the contarct that can be represented by ANTLR grammar, \cite{Parr13}. The high level structure is presented in Listing \ref{ctrgrammar} and mainly follows that of Java grammar provided by ANTLR.

\begin{lstlisting} [caption={Contract Grammar},label=ctrgrammar]
parser grammar ContractParser;

options {
    tokenVocab = ContractLexer;
}

// Parser Rules

program
     : contractDeclaration EOF
     ;
contractDeclaration
     : CONTRACT identifier 
     OF identifier
     contractBody
     ;

contractBody
     : '{' contractBodyDeclaration* '}'
     ;
contractBodyDeclaration
     :';'
     | STATIC? block
     | modifier* itemDeclaration
     ;

itemDeclaration
     : contrmethodDeclaration
     | fieldDeclaration
     | protocolDeclaration
     ;

protocolDeclaration
     : PROTOCOL '{' grammarString+ '}'
     ;
grammarString
     : IDENTIFIER | '(' | ')' | ADD | MUL | SEMI | CARET | '|' | '?'
     ;


contrmethodDeclaration
    : typeTypeOrVoid identifier formalParameters ('[' ']')* (THROWS qualifiedNameList)? methodBody
    ;
\end{lstlisting}

\subsection{Parsing Contracts to Graph}
\label{sec:parsing}
\begin{figure}[h]
\centering
\includegraphics[width=3.00in,height=2.25in]{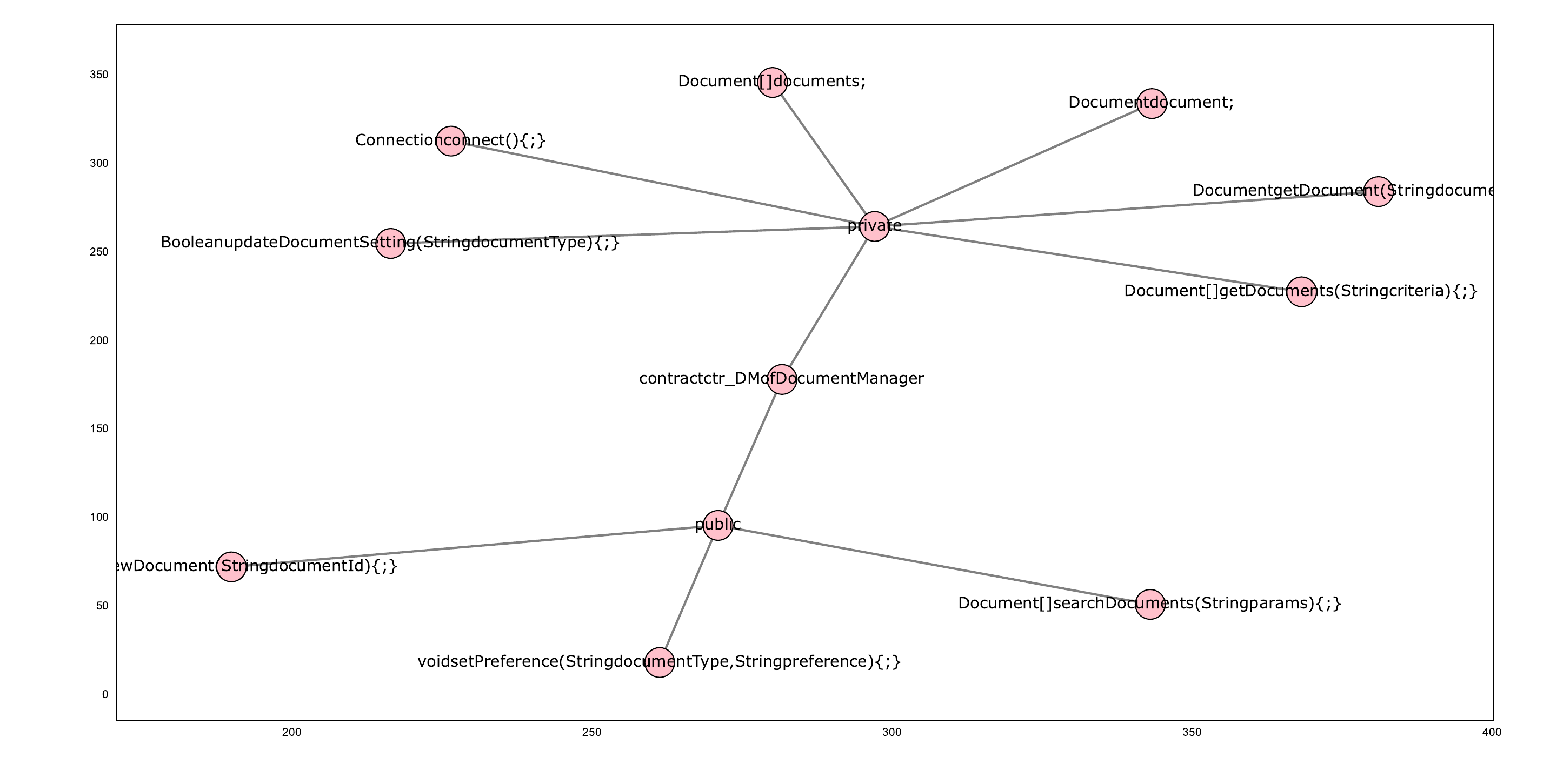}
\caption{Graph Representation of the Contract - Document Manager.}
\label{fig:contractgraph}
\end{figure}

ANTLR Tools are used to parse the contract files containing the contract specification that satisfies the grammar specification in Listing \ref{ctrgrammar}. We have written additional code to convert the contract into a graph (based on the NETOWRKX libraries in Python). The graph generated for the Listing shown in Listing \ref{contrDM} shown in Figure \ref{fig:contractgraph}.

\subsection{Creating the Application Architecture Graphs}
\label{sec:graph}

Algorithm \ref{alg:aagraph} describes the pseudo code to build the application architecture graph.

\begin{algorithm}
\caption{Building Application Architecture Graph}\label{alg:aagraph}
	\begin{algorithmic}[1]
	\Require {Contract Specification Files for all Components, $ctr \in \{ contacrts$\}}
	\State Crate the Application Architecture Graph root node as ``''AA''' using the $networkx$ package
	\For {Every $ctr \in \{contract\}$}
	\State Parse the ctr file to geneate the graph for the contarct with the root node ctr.name (name of the contrtact)
	\State Cretae and edge between ``AA'' and ctr.name
	\EndFor
	\State Save the generated graph in graphml format

	\end{algorithmic} 
\end{algorithm} 
GraphML is a comprehensive and easy-to-use file format for graphs\footnote{\url{http://graphml.graphdrawing.org/}} is used to save the graphs.

\section{Matching Implementation}
\label{sec:matching}
\subsection{Subgraph Matching}\label{sec:subgraphmatch}
We are looking to solve the graph isomorphism problem where we are tryinh to macth the required business contract with a subgraph of tha application architecture graph. This problem is known to be NP-complete, \cite{10.1145/800157.805047}. There are several algorithms to solve this problem, see \cite{1323804}, which proposes the VF2 algorithm that has been implemented by the networkx python library we are using in this framework.

\begin{definition}
Graph Isomorphism: Two graphs are isomorphic if they have the same number of edges, vertices, and same edges connectivity. Formally graphs G and H are isomorphic if we can establish a bijection between the vertex sets of G and H. $f : N(G) \rightarrow N(G)$, such as if  v and  w are adjacent in $G \Rightarrow$ f(v) and f(w) are adjacent in H.
\end{definition}

\subsection{Protocol Equivalence}\label{sec:protocolmatch}
Protocols are expressed as regular expressions. In order to compare two protocols, we first convert these to minimal DFA (deterministic finite automata) and then compare the two see if they are equivalent see \cite{10.5555/524279}.

\begin{figure}[h]
\centering
\includegraphics[width=3.00in,height=1.301in]{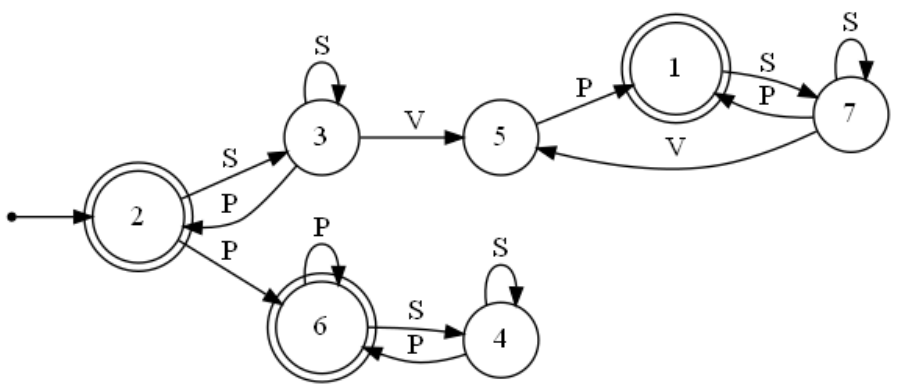}
\caption{Protocol of Contract DM as DFA}
\label{fig:dfa}
\end{figure}

Fig.\ref{fig:dfa} shows the protocol of the contract shown in Listing \ref{contrDM} as a DFA. We have shortened the names where we have set ``searchDocument=D'', ``setPreference=P'' and ``viewDocument=V''.

\subsection{Two phase approach}
\begin{figure}[h]
\centering
\includegraphics[width=3.00in,height=3.4948in]{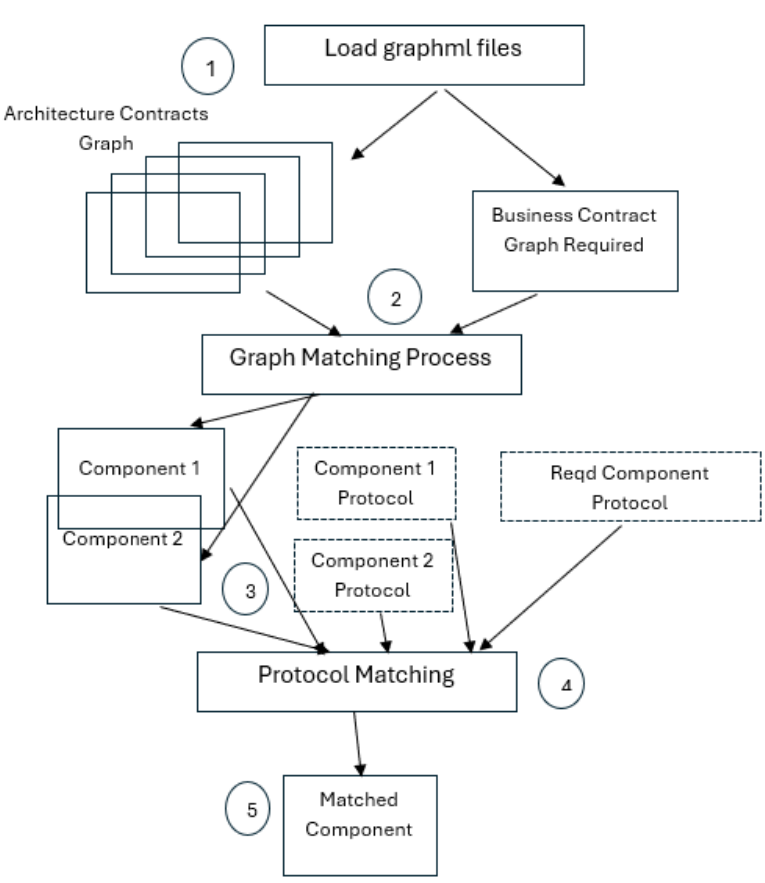}
\caption{Matching Process - Graph and Protocol Matching: In stap (1) We load the contract graphs stored as graphml files. (2) We create the application architecture graph as well as the required component graph an apply the subgraph matching algorithm to generate the matching contracts shown in step (3). We then extract the protocols from the matched components and apply the protocol matching algorithm to step (4). Finally we get the matched component as a result in step (5).}
\label{fig:bcap}
\end{figure}

We implement a two-phased approach to solve the component matching problem, see Fig.\ref{fig:bcap}.In the first phase we load the contract graph files, then we match the required business architecture component graph with the application architecture contract graph to generate a subset of the components. If this is empty, then we do not have a match the recommendation will be to build the new component.

In the case we find matches, we then extract the protocols from these components and then match it with the required protocol.  If so, then we have one or more matches, and if not we again recommend that we build the new component. 

\section{Conclusions}
\label{sec:conclusion}
In this paper we addressed the problems of automatically matching and identifying software components for a required component from the set of components of the Application Architecture. We used contract specification to represent the component. A grammar was developed so that we can process these contracts. In our approach these contracts are converted to graphs and can be storeed as graphml files. We use protocol to represent the interatcions of a component and the matching process is implementes as a two-phase approach. In the first phase we do a subgraph matching to identfy a set of components that match the required contract. In the second phas we match the protocols to refine the match. We have used very efficient algorithms available in the Python networks library to implement the first phase and use the approach of converting the protocol (regular expression) to DFAs and mactching the DFAs.





\bibliographystyle{elsarticle-num-names}
\bibliography{capability}







\end{document}